\documentclass[twocolumn,tighten]{aastex62}
\def\kms{{km s$^{-1}$}}

\shorttitle{Dynamical Coherence in Several-Mpc Scales}
\shortauthors{Lee et al.}
\def\simlt{\lower.5ex\hbox{$\; \buildrel < \over \sim \;$}}
\def\simgt{\lower.5ex\hbox{$\; \buildrel > \over \sim \;$}}

\begin{document}

\title{Mysterious Coherence in Several-Megaparsec Scales Between Galaxy Rotation and Neighbor Motion}

\author{Joon Hyeop Lee}
\email{jhl@kasi.re.kr}
\author{Mina Pak}
\affil{Korea Astronomy and Space Science Institute, Daejeon 34055, Republic of Korea}
\affil{University of Science and Technology, Daejeon 34113, Republic of Korea}

\author{Hyunmi Song}
\affil{Korea Astronomy and Space Science Institute, Daejeon 34055, Republic of Korea}

\author{Hye-Ran Lee}
\affil{Korea Astronomy and Space Science Institute, Daejeon 34055, Republic of Korea}
\affil{University of Science and Technology, Daejeon 34113, Republic of Korea}

\author{Suk Kim}
\affil{Korea Astronomy and Space Science Institute, Daejeon 34055, Republic of Korea}
\affil{Center for Galaxy Evolution Research, Yonsei University, Seoul 03722, Republic of Korea}
\affil{Department of Astronomy and Space Science, Chungnam National University, Daejeon 34134, Republic of Korea}

\author{Hyunjin Jeong}
\affil{Korea Astronomy and Space Science Institute, Daejeon 34055, Republic of Korea}

\begin{abstract}
In our recent report, observational evidence supports that the rotational direction of a galaxy tends to be coherent with the average motion of its nearby neighbors within 1 Mpc. We extend the investigation to neighbors at farther distances, in order to examine if such dynamical coherence is found even in large scales. The Calar Alto Legacy Integral Field Area (CALIFA) survey data and the NASA-Sloan Atlas (NSA) catalog are used. From the composite map of velocity distribution of `neighbor' galaxies within 15 Mpc from the CALIFA galaxies, the composite radial profiles of the luminosity-weighted mean velocity of neighbors are derived. These profiles show unexpectedly strong evidence of the dynamical coherence between the rotation of the CALIFA galaxies and the average line-of-sight motion of their neighbors within several Mpc distances.
Such a signal is particularly strong when the neighbors are limited to red ones: the luminosity-weighted mean velocity at $1<D\le6$~Mpc is as large as $30.6\pm10.9$ {\kms} ($2.8\sigma$ significance to random spin-axis uncertainty) for central rotation ($R\leq R_e$).
In the comparison of several subsamples, the dynamical coherence tends to be marginally stronger for the diffuse or kinematically-well-aligned CALIFA galaxies. For this mysterious coherence in large scales, we cautiously suggest a scenario that it results from a possible relationship between the long-term motion of a large-scale structure and the rotations of galaxies in it.
\end{abstract}


\section{INTRODUCTION}\label{intro}

Galaxy kinematics provides important clues to trace the formation history of a galaxy.
Particularly, galaxy rotation is a simple but strong constraint on the past events of galaxy assembly, because angular momentum is always conserved in an isolated system. If a galaxy formed from a rotating gas cloud, the angular momentum of the gas cloud must remain in the galaxy after condensation. If a galaxy formed from an off-axis merger of two objects, the total angular momentum of the binary system must be succeeded by the merger-remnant galaxy. This is simple physics, but in reality the detailed origins of galaxy rotation are not sufficiently understood yet. This is partially because the history of integral field spectroscopy (IFS) is not so long and thus until only several years ago it was not easy to secure a galaxy sample that is large enough to obtain statistically reliable results.

In the last decade, however, we have learned various aspects about how galaxy rotation is influenced by environment. Owing to several large IFS surveys, now it is known that even early-type galaxies mostly rotate and such rotation is tightly related to environmental density \citep[][and many other studies]{cap06,ems07, cap11,ems11,kra11}. It was also revealed that direct interactions or mergers between galaxies significantly affect the position angle of galaxy rotation axis, which may result in prolate rotation \citep{tsa17,kra18,wea18}, morpho-kinematic misalignment \citep{bar15,oh16}, or kinematically distinct cores \citep{ems14,kra15,tay18}.

\begin{figure}[t]
\centering
\plotone{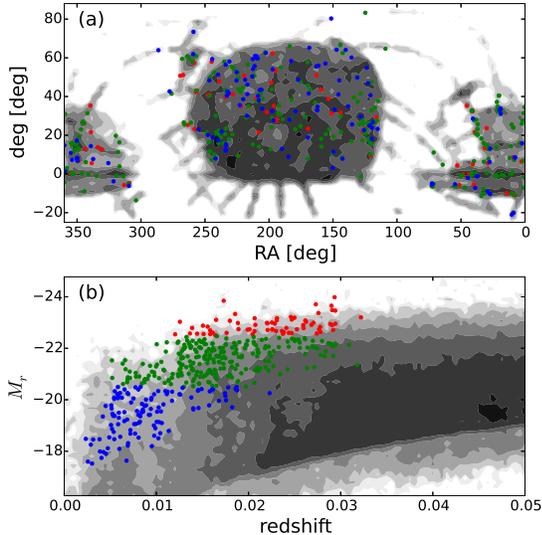}
\caption{(a) Spatial distribution and (b) redshits and $r$-band absolute magnitudes of the CALIFA galaxies (colored dots: red for $M_r\leq -22.5$, green for $-22.5<M_r\leq -20.5$ and blue for $M_r> -20.5$). The background contours show the log-scale number density of the NSA galaxies.\label{spdist}}
\end{figure}

More recently, \citet[][hereafter \emph{L19}]{lee19} reported the first discovery of observational evidence for the systematic coherence between galaxy rotation and the average motion of neighbor galaxies. Such coherence appears to be particularly strong for the rotation at the outskirt ($R_e<R\leq 2R_e$; where $R_e$ is the effective radius) of each galaxy. The coherence signal is statistically significant for neighbors out to 800 kpc from each target galaxy, and it tends to be more conspicuous when target galaxies are faint and neighbor galaxies are bright. All of these results indicate that fly-by interactions with neighbors may strongly influence the rotational direction of a galaxy at least in its outskirt.

In \emph{L19}, the luminosity-weighted mean velocity profiles of neighbors significantly drop down at $\sim800$ kpc and almost converge to zero velocity as the distance from a target galaxy increases (Figure~13 in \emph{L19}), which seems to indicate that too distant neighbors hardly affect galaxy rotation, as we can reasonably guess. However, the profiles appear to be somewhat noisy and fluctuating, and thus one may have suspicion if the coherence signals really converge to zero at $>800$ kpc. This suspicion can be rephrased as the following question: is galaxy rotation not related to the motions of neighbors at far distances (for example, in several-Mpc scales) at all?

As several recent observational and theoretical studies have reported, the spin axes of galaxies appear to be aligned with the directions of surrounding large-scale structures such as filaments, which means that large-scale structures may influence the internal kinematics of individual galaxies to some extent \citep[e.g.,][]{nav04,tem13,lai15,kim18,lee18}. %,jeo19}.
In that viewpoint, will it be possible that some large-scale effects cause dynamical coherence between galaxies at far distances, although the direct interactions between them are impossible?

To answer these questions, in this paper, we extend the previous work of \emph{L19} to larger scales, out to 15 Mpc.
The paper follows the listed structure. Section~\ref{data} describes the data set and key quantities. Section~\ref{anal} specifies our methods to detect the signal of the dynamical coherence in large scales. Section~\ref{result} shows the results, and a possible scenario for the results is discussed in Section~\ref{discuss}. The conclusions of the paper are given in Section~\ref{conclude}.
Throughout this paper, we adopt the cosmological parameters: $h=0.7$, $\Omega_{\Lambda}=0.7$, and $\Omega_{M}=0.3$.

\section{DATA, SAMPLE AND QUANTITIES}\label{data}

\begin{figure}[t]
\centering
\plotone{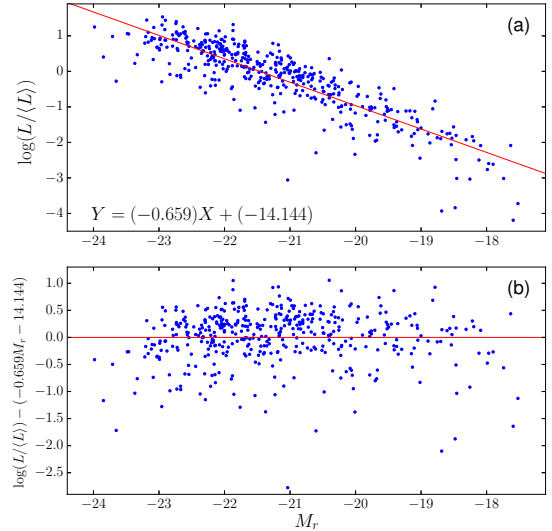}
\caption{(a) Normalized angular momenta ($R\leq R_e$) of the CALIFA galaxies as a function of $r$-band absolute magnitude. The red line is the linear regression fit. (b) Normalized angular momenta ($R\leq R_e$), corrected for $r$-band absolute magnitude.  \label{angmom}}
\end{figure}

In this paper, we use the PyCASSO database\footnote{http://pycasso.ufsc.br or http://pycasso.iaa.es/} \citep{dea17}, which is a data set value-added by analyzing the Calar Alto Legacy Integral Field Area Survey \citep[CALIFA;][]{san12,san16,wal14} data with the Python CALIFA STARLIGHT Synthesis Organizer platform \citep[PyCASSO;][]{cid05,cid13}. The PyCASSO database provides well-produced maps of various spectroscopic information including stellar mass and line-of-sight velocity for 445 galaxies at $z\lesssim0.03$. The greatest merit of CALIFA is that it has unprecedentedly wide field of view ($> 1$ arcmin$^2$), which covers more than $2R_e$ for most targets. For more detailed description about the CALIFA and PyCASSO data, see Section~2.1 of \emph{L19} and the references therein.
Figure~\ref{spdist} presents the sky distribution of the CALIFA galaxies and their absolute magnitudes and redshifts. The CALIFA galaxies are not significantly biased or clustered to any specific region in the sky.

From the PyCASSO maps of stellar mass surface density and line-of-sight velocity of each CALIFA galaxy, we estimated the angular momenta at its center ($R\leq R_e$) and outskirt ($R_e<R\leq 2R_e$). We also estimated the statistical uncertainty of the position angle for each angular momentum vector, by bootstrapping spaxels in each target galaxy, as described in Section~2.1.1 of \emph{L19}. In this paper, we limit our sample to the CALIFA galaxies that have at least five Voronoi bins at given radial range and the position angle uncertainty not larger than $45^{\circ}$, which leaves 434 galaxies with the central angular momentum measurements and 392 galaxies with the outskirt angular momentum measurements.

Figure~\ref{angmom}(a) shows the central angular momenta normalized by the mean value among the CALIFA galaxies, as a function of $r$-band absolute magnitude. Since the angular momenta strongly depend on the absolute magnitudes, largely due to the mass factor in the angular momentum formula ($L=mr\times v$), we derived the \emph{corrected angular momenta} to remove their luminosity dependence, as shown in Figure~\ref{angmom}(b). This correction enables us to simply separate between the galaxies with low and high angular momenta in any luminosity bin.

\begin{figure}[t]
\centering
\plotone{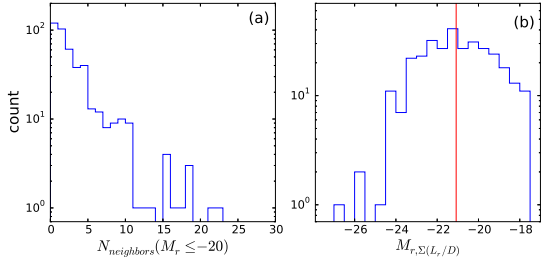}
\caption{Histograms of the indicators for local environments around the CALIFA galaxies: (a) the number of neighbors brighter than $M_r=-20$ in 1 Mpc and $\pm500$ {\kms}, and (b) the local luminosity density, which is defined in the main text (a smaller value indicates higher density). The median value of the local luminosity density of the CALIFA galaxies with at least one neighbor is denoted by the red line ($M_{r,\Sigma(L_r/D)}=-21.1$).  \label{env}}
\end{figure}

\begin{figure}[t]
\centering
\plotone{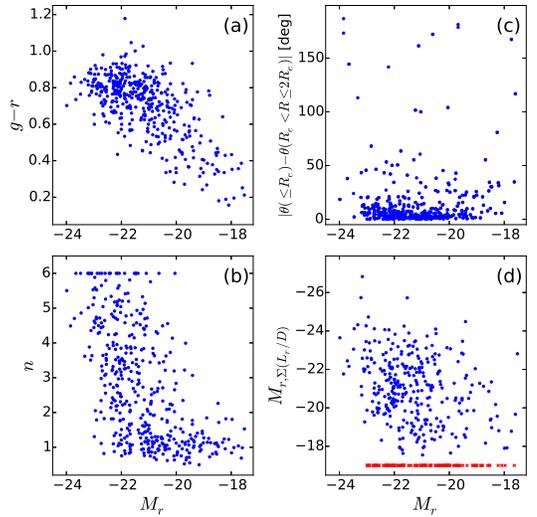}
\caption{Several basic quantities of the CALIFA galaxies as a function of $r$-band absolute magnitude: (a) $g-r$ color, (b) S{\'e}rsic index, (c) internal angular misalignment, and (d) local luminosity density. The red crosses are the CALIFA galaxies that have no neighbor ($M_r \leq -20$) within 1 Mpc distance. \label{basic}}
\end{figure}

For the estimation of the neighbor motions around the CALIFA galaxies, we use the NASA-Sloan Atlas (NSA) catalog\footnote{http://www.nsatlas.org}. The NSA catalog was created by Michael Blanton, by combining the Sloan Digital Sky Survey \citep[SDSS;][]{yor00}, NASA Extragalactic Database (NED)\footnote{https://ned.ipac.caltech.edu/}, Six-degree Field Galaxy Redshift Survey \citep[6dFGS;][]{jon09}, Two-degree Field Galaxy Redshift Survey \citep[2dFGRS;][]{col01}, CfA Redshift Survey \citep[ZCAT;][]{huc83}, Arecibo Legacy Fast ALFA Survey \citep[ALFALFA;][]{gio05} and the Galaxy Evolution Explorer \citep[GALEX;][]{mar03} survey data.
From the NSA catalog, we obtained right ascension, declination, redshift, S{\'e}rsic index and absolute magnitudes in the $g$ and $r$ bands for the CALIFA galaxies and their neighbors.

We define the `neighbors' as the galaxies that have line-of-sight velocity differences within $\pm500$ {\kms} and projected distances not larger than 15 Mpc from the CALIFA galaxies. Note that the galaxies at such huge distances are not usually called `neighbors', but as did in \emph{L19}, we keep this wording for convenience in this paper.

To define the local environment of each CALIFA galaxy, we first counted the number of neighbors that satisfy: (1) the $r$-band absolute magnitude is not fainter than  $M_r= -20$ \footnote{This magnitude cut is applied only to the local luminosity density calculation. No magnitude cut is applied to the calculation of luminosity-weighted mean velocity profiles.}, (2) the distance from the CALIFA galaxy is not larger than 1 Mpc, and (3) the line-of-sight velocity difference is not larger than 500 {\kms}. Among the 434 CALIFA galaxies, 122 galaxies do not have any neighbor that satisfies the conditions.
For the CALIFA galaxies that have one or more neighbors, we estimated the luminosity density weighted by distance ($M_{r,\sum(L_r/D)}$), as follows:
\begin{equation}
 M_{r,\sum(L_r/D)} = -2.5 \log \Big(\sum^{i} 10^{-0.4M_{r,i}}/D_i \Big),
\end{equation}
where $M_{r,i}$ is the $r$-band absolute magnitude of the $i$-th neighbor and $D_i$ is its distance from a given CALIFA galaxy in unit of 100 kpc. This parameter is a rough proxy of the integrated gravitational potential from the neighbors \citep[the basic concept is introduced in][]{lee16}. 
%For example, if a CALIFA galaxy has a single neighbor with $M_r=-21$ at the distance of 100 kpc, the luminosity density is $M_{r,\sum(L_r/D)} = -21$, which is equivalent with the case that two neighbors with $M_r=-21$ at the distance of 200 kpc.
In Figure~\ref{env}, it is shown how the number of neighbors and the local luminosity density of the CALIFA galaxies are distributed. The median value of the luminosity density of the CALIFA galaxies with at least one neighbor is $M_{r,\sum(L_r/D)} = -21.1$.

Figure~\ref{basic} presents the distributions of several quantities as a function of $r$-band absolute magnitude for the CALIFA galaxies. The $g-r$ color appears to strongly depend on magnitude as well known, while the S{\'e}rsic index and local luminosity density show weak dependence. The internal angular misalignment, defined as the position angle difference between the central and outskirt angular momentum vectors ($|\theta(\leq R_e) - \theta(R_e<R\leq 2R_e)|$), hardly depends on magnitude, as described in \emph{L19}.

\section{ANALYSIS}\label{anal}

\begin{figure}[t]
\centering
\plotone{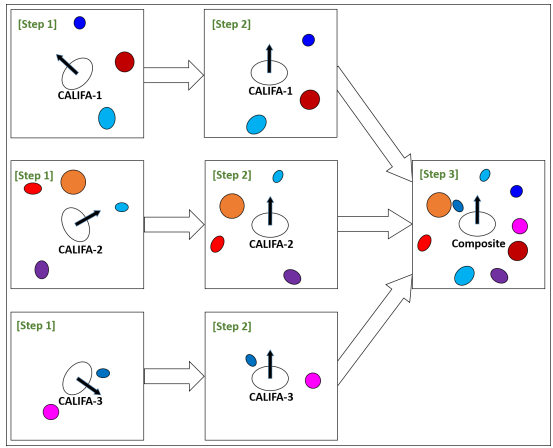}
\caption{Schematic picture describing the procedure to build a composite kinematics map of the neighbors around the CALIFA galaxies. Step 1 (left boxes): kinematics maps for individual systems. Step 2 (middle boxes): kinematics maps aligned for the angular momentum vector of each CALIFA galaxy to be upward. Step 3 (right box): the composite map of kinematics for the whole systems around the CALIFA galaxies. \label{flow}}
\end{figure}

The procedure to investigate the dynamical coherence in large scales is intrinsically the same as the work in small scales of \emph{L19}. The final goal of the procedure is to build the luminosity-weighted mean velocity profiles with statistical uncertainties, from which we can determine how significant the coherence between galaxy rotation and the average motion of neighbors at given distance is.
Since the details of the procedure are fully described in Sections~3 and 4 of \emph{L19}, here we simply summarize the key processes from the individual kinematics maps to the luminosity-weighted mean velocity profiles with statistical uncertainties.

\begin{figure}[t]
\centering
\plotone{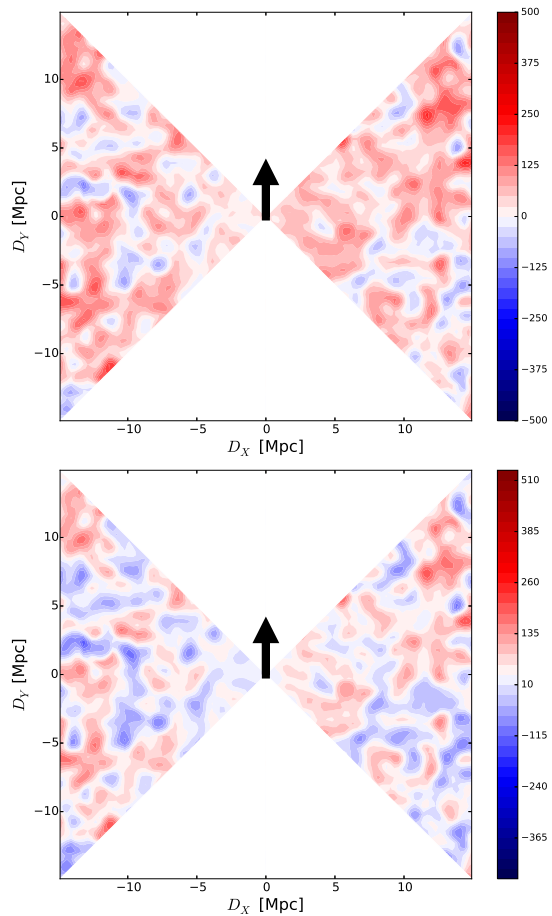}
\caption{\emph{Upper}: Contour map for the luminosity-weighted mean line-of-sight velocity of neighbors out to 15 Mpc, estimated using the composite kinematics map (aligned for the central rotations of the CALIFA galaxies). The right-side bar shows the color code, in which the numbers indicate the line-of-sight velocities in unit of {\kms}. The contour map is built on 100 $\times$ 100 bins, and each bin was smoothed over 3 bins with linear weight by distance. Note that this smoothing is just for the visualization and irrelevant to the main results (the mean velocity profiles). \emph{Lower}: The same as the upper panel, but the color bar is shifted by 35 {\kms} (the mean value of the cumulative luminosity-weighted mean velocities at $D\sim15$ and $-15$ Mpc; see Figure~\ref{vprof}(b) and the main text) to compensate the redshift bias of the CALIFA galaxies. \label{velmap}}
\end{figure}

\begin{figure}[t]
\centering
\plotone{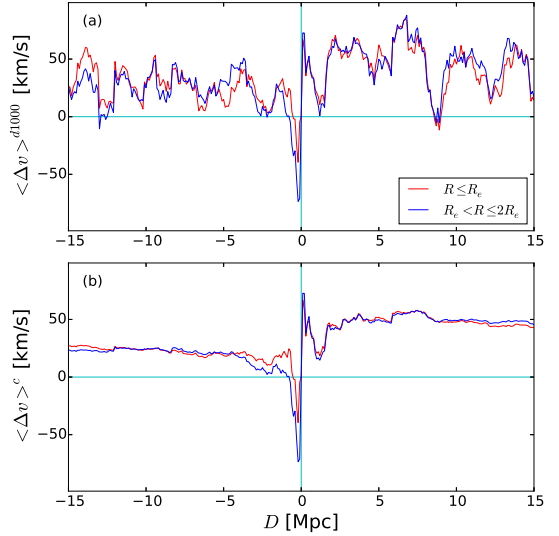}
\caption{(a) Derivative radial profiles of luminosity-weighted mean velocity. A top-hat smoothing kernel with 1-Mpc size is applied. The red line shows the profiles for central angular momenta ($R{\le}R_e$), while the blue line is the profiles for outskirt angular momenta ($R_e<R\le 2R_e$). (b) Cumulative radial profiles of luminosity-weighted mean velocity. The positive/negative values in the distance from a given CALIFA galaxy ($D$) indicate the right/left-side neighbors. \label{vprof}}
\end{figure}

\begin{enumerate}
 \item[(1)] First of all, a composite kinematics map needs to be built from the kinematics maps for individual CALIFA galaxies and their neighbors, because the number of neighbors in a single system is mostly not enough to give sufficient reliability in the statistical analysis. This process is schematized in Figure~\ref{flow}. The individual systems are aligned for the angular momentum vector of each CALIFA galaxy to be upward (Step~2 in Figure~\ref{flow}). After that, the whole systems are combined into a single composite map of kinematics (Step~3 in Figure~\ref{flow}).
 \item[(2)] In the composite map, the neighbors in the domains of $-45^{\circ}<\theta<45^{\circ}$ and $135^{\circ}<\theta<225^{\circ}$ are discarded, where $\theta$ is the position angle from the angular momentum vector direction (this configuration is called \emph{X-cut}; \emph{L19}), in consideration of the uncertainty in measuring the position angle of an angular momentum vector. Figure~\ref{velmap} shows the luminosity-weighted line-of-sight velocity contour maps after the X-cut. If the dynamical coherence exists in large scales, the right-side contours must be redder than the left-side contours on average.
 \item[(3)] We estimate the luminosity-weighted mean velocity profiles, as shown in Figure~\ref{vprof}. The derivative mean velocity ($\langle\Delta v\rangle^{d1000}$) is defined as follows:
\end{enumerate}
\begin{equation}
 \langle\Delta v\rangle^{d1000}(D') = \left\{ 
 \begin{array}{ll}
   \frac{\displaystyle \sum_{Rd(D',1000)} \Delta v \mathcal{L}}{\displaystyle \sum_{Rd(D',1000)} \mathcal{L}} & \textrm{if}\: D'>0 \\
   0 & \textrm{if}\: D'=0 \\
   \frac{\displaystyle \sum_{Ld(D',1000)} \Delta v \mathcal{L}}{\displaystyle \sum_{Ld(D',1000)} \mathcal{L}} & \textrm{if}\: D'<0 \textrm{,}
 \end{array} \right .
\end{equation}
where $\Delta v$ is the line-of-sight recession velocity of a neighbor galaxy relative to a given CALIFA galaxy, $\mathcal{L}$ is the luminosity of the neighbor galaxy, $D'$ is the projected distance to the CALIFA galaxy, and the right-side distance range $Rd$ is:
\begin{equation}
 Rd(D',1000) = \left\{ 
 \begin{array}{l}
   D'-1\,\textrm{Mpc} < D \le D' \\
   \qquad\qquad\, \textrm{if}\: D'>1 \,\textrm{Mpc}\\
   0 < D \le D' \\
   \qquad\qquad\, \textrm{if}\: 0<D'\le 1\,\textrm{Mpc,}
 \end{array} \right .
\end{equation}
and the left-side distance range $Ld$ is:
\begin{equation}
 Ld(D',1000) = \left\{ 
 \begin{array}{l}
   D' \le D < D'+1\,\textrm{Mpc} \\
   \qquad\qquad \textrm{if}\: D'<-1 \,\textrm{Mpc}\\
   D' \le D < 0 \\
   \qquad\qquad \textrm{if}\: -1\,\textrm{Mpc}\le D'<0 \textrm{,}
 \end{array} \right .
\end{equation}
and the cumulative mean velocity ($\langle\Delta v\rangle^{c}$) is:
\begin{equation}
 \langle\Delta v\rangle^{c}(D') = \left\{ 
 \begin{array}{ll}
   \frac{\displaystyle \sum_{0<D\le D'} \Delta v \mathcal{L}}{\displaystyle \sum_{0<D\le D'} \mathcal{L}} & \textrm{if}\: D'>0 \\
   0 & \textrm{if}\: D'=0 \\
   \frac{\displaystyle \sum_{D'\le D<0} \Delta v \mathcal{L}}{\displaystyle \sum_{D'\le D<0} \mathcal{L}} & \textrm{if}\: D'<0 \textrm{.}
 \end{array} \right .
\end{equation}

\begin{figure*}[t]
\centering
\includegraphics[width=0.95\textwidth]{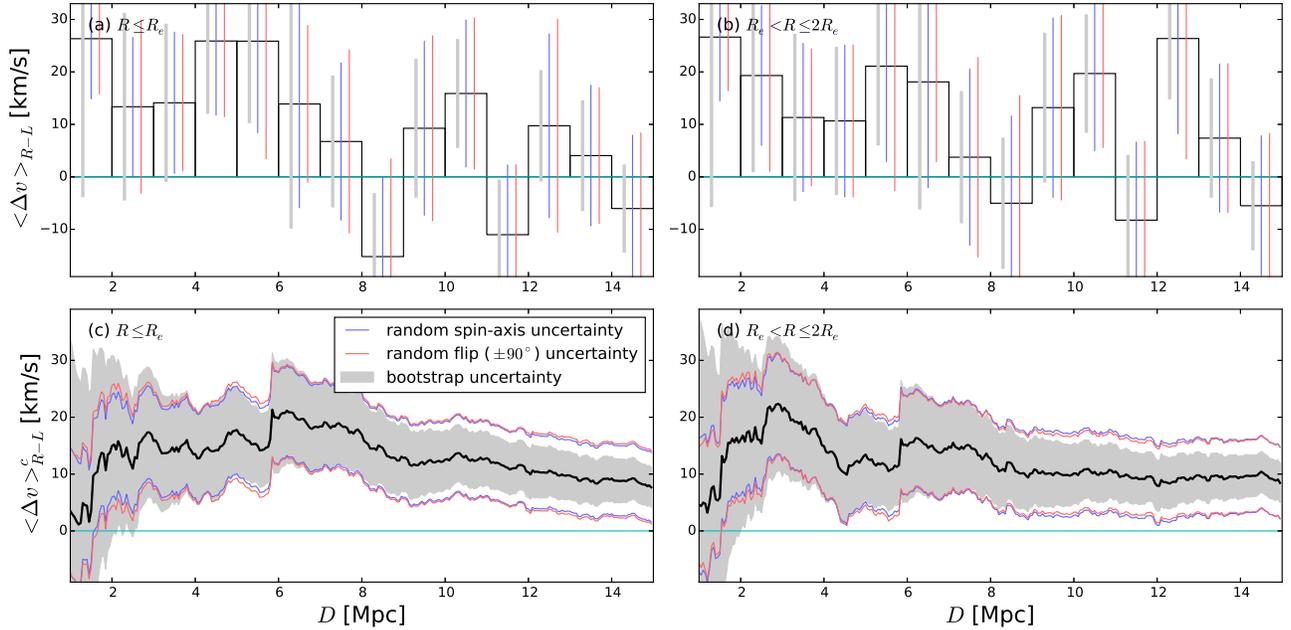}
\caption{Right-left-merged luminosity-weighted mean velocity profiles for the whole CALIFA sample: (a) the 1-Mpc-binned mean velocity distribution for central ($R\leq R_e$) rotation, (b) the 1-Mpc-binned mean velocity distribution for outskirt ($R_e<R\leq 2R_e$) rotation, (c) the cumulative profile for central rotation, and (d) the cumulative profile for outskirt rotation. Three different kinds of statistical uncertainty are denoted: bootstrap uncertainty (BST; shades), random spin-axis uncertainty (RAX; blue lines), and randomly flipped ($\pm90^{\circ}$) spin-axis uncertainty (RFA; red lines). \label{vprofall}}
\end{figure*}

\begin{figure*}[p]
\centering
\includegraphics[width=0.95\textwidth]{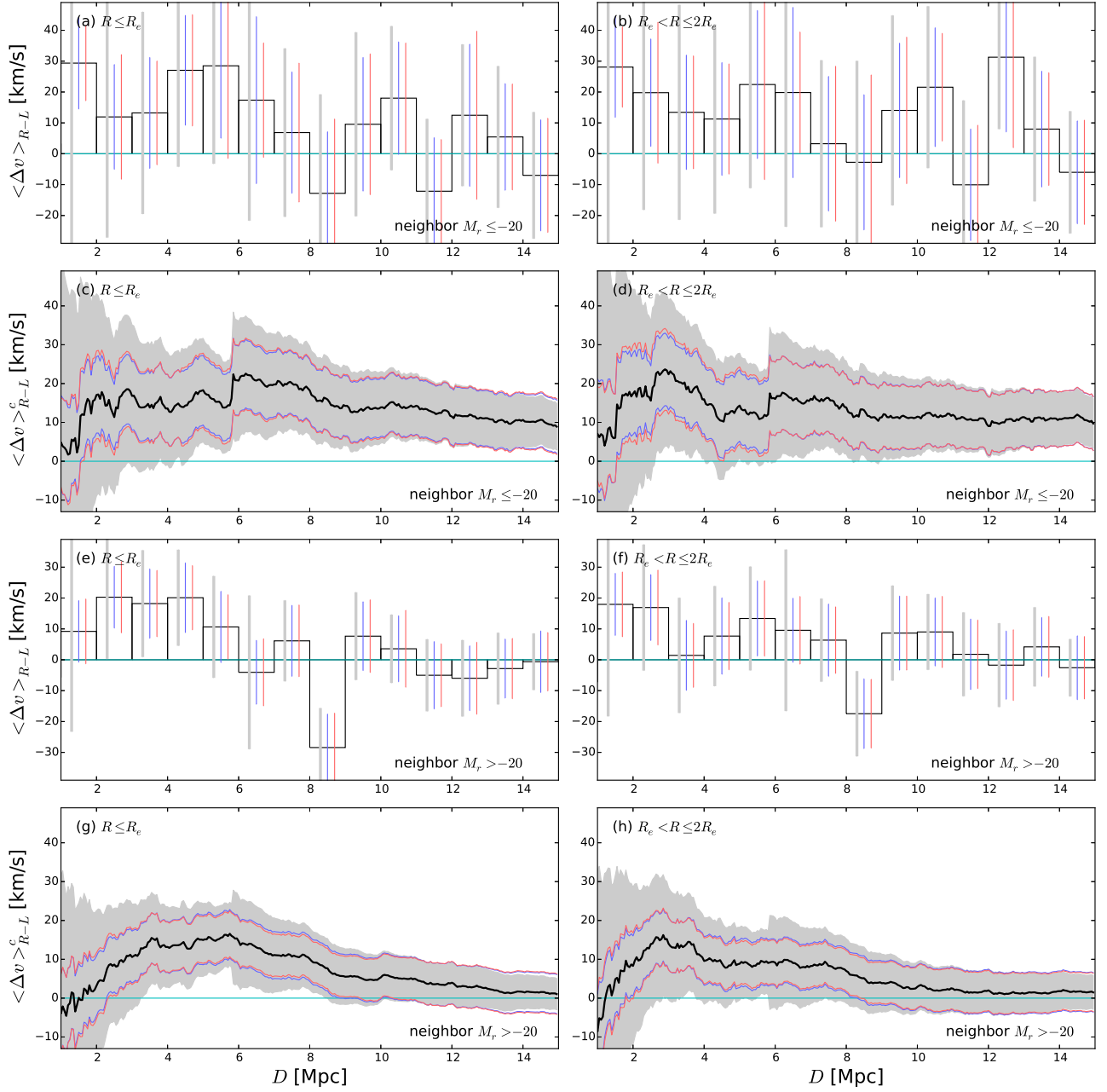}
\caption{Right-left-merged luminosity-weighted mean velocity profiles for the whole CALIFA galaxies, when the neighbors are divided by luminosity: (a) the 1-Mpc-binned mean velocity distribution for central rotation and for bright ($M_r\leq -20$) neighbors, (b) the 1-Mpc-binned mean velocity distribution for outskirt rotation and for bright neighbors, (c) the cumulative mean velocity profile for central rotation and  for bright neighbors, (d) the cumulative mean velocity profile for outskirt rotation and  for bright neighbors, (e) the 1-Mpc-binned mean velocity distribution for central rotation and for faint ($M_r>-20$) neighbors, (f) the 1-Mpc-binned mean velocity distribution for outskirt rotation and for faint neighbors, (g) the cumulative mean velocity profile for central rotation and  for faint neighbors, and (h) the cumulative mean velocity profile for outskirt rotation and  for faint neighbors. \label{vprofneimag}}
\end{figure*}

\begin{figure*}[p]
\centering
\includegraphics[width=0.95\textwidth]{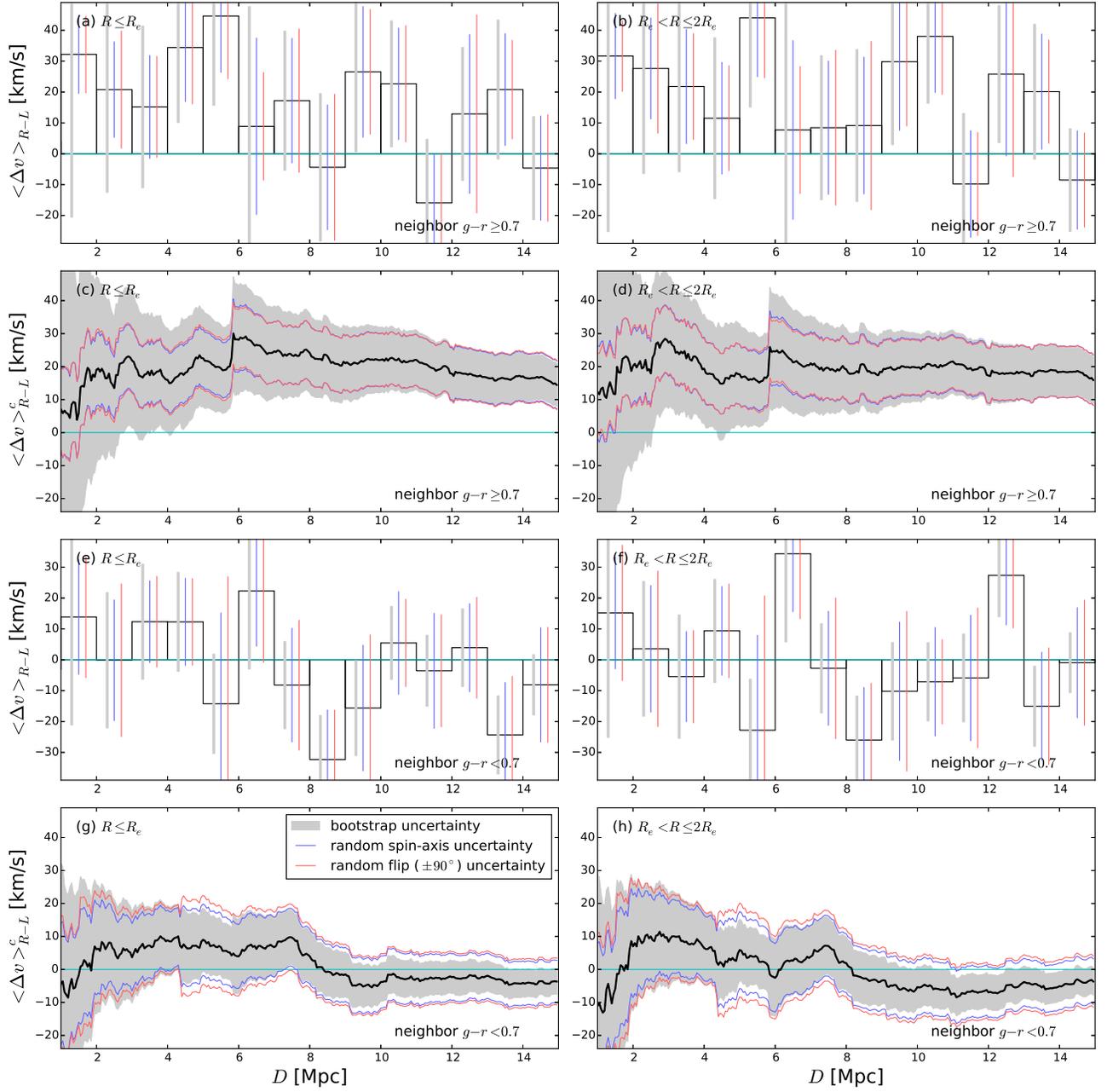}
\caption{Right-left-merged luminosity-weighted mean velocity profiles for the whole CALIFA galaxies, when the neighbors are divided by color: (a) the 1-Mpc-binned mean velocity distribution for central rotation and for red ($g-r\geq 0.7$) neighbors, (b) the 1-Mpc-binned mean velocity distribution for outskirt rotation and for red neighbors, (c) the cumulative mean velocity profile for central rotation and  for red neighbors, (d) the cumulative mean velocity profile for outskirt rotation and  for red neighbors, (e) the 1-Mpc-binned mean velocity distribution for central rotation and for blue ($g-r<0.7$) neighbors, (f) the 1-Mpc-binned mean velocity distribution for outskirt rotation and for blue neighbors, (g) the cumulative mean velocity profile for central rotation and  for blue neighbors, and (h) the cumulative mean velocity profile for outskirt rotation and  for blue neighbors. \label{vprofneicol}}
\end{figure*}

\begin{enumerate}
 \item[(4)] The positive velocities of the right-side neighbors ($D>0$) and the negative velocities of the left-side neighbors ($D<0$) commonly support the coherence between galaxy rotation and neighbor motions. Therefore, we can further simplify Figure~\ref{vprof}(b) by defining the \emph{right-left-merged} mean velocities (Figures~\ref{vprofall} - \ref{vprofneicol}), as follows:
\end{enumerate}
\begin{equation}\label{cumeqn}
 \langle\Delta v\rangle^{c}_{R-L}(D') = \frac{\displaystyle \Bigg(\sum_{0<D{\le}D'} \Delta v \mathcal{L} \Bigg) - \Bigg(\sum_{-D'{\le}D<0} \Delta v \mathcal{L} \Bigg)}{\displaystyle \Bigg( \sum_{0<D{\le}D'} \mathcal{L} \Bigg) + \Bigg(\sum_{-D'{\le}D<0} \mathcal{L} \Bigg)}
\end{equation}
where $D'>0$. While Equation~\ref{cumeqn} defines the cumulative mean velocity profile ($\langle\Delta v\rangle^{c}_{R-L}$), the right-left-merged mean velocity at any given distance range without accumulation is simply denoted as $\langle\Delta v\rangle_{R-L}$.

\begin{figure*}[t]
\centering
\includegraphics[width=0.95\textwidth]{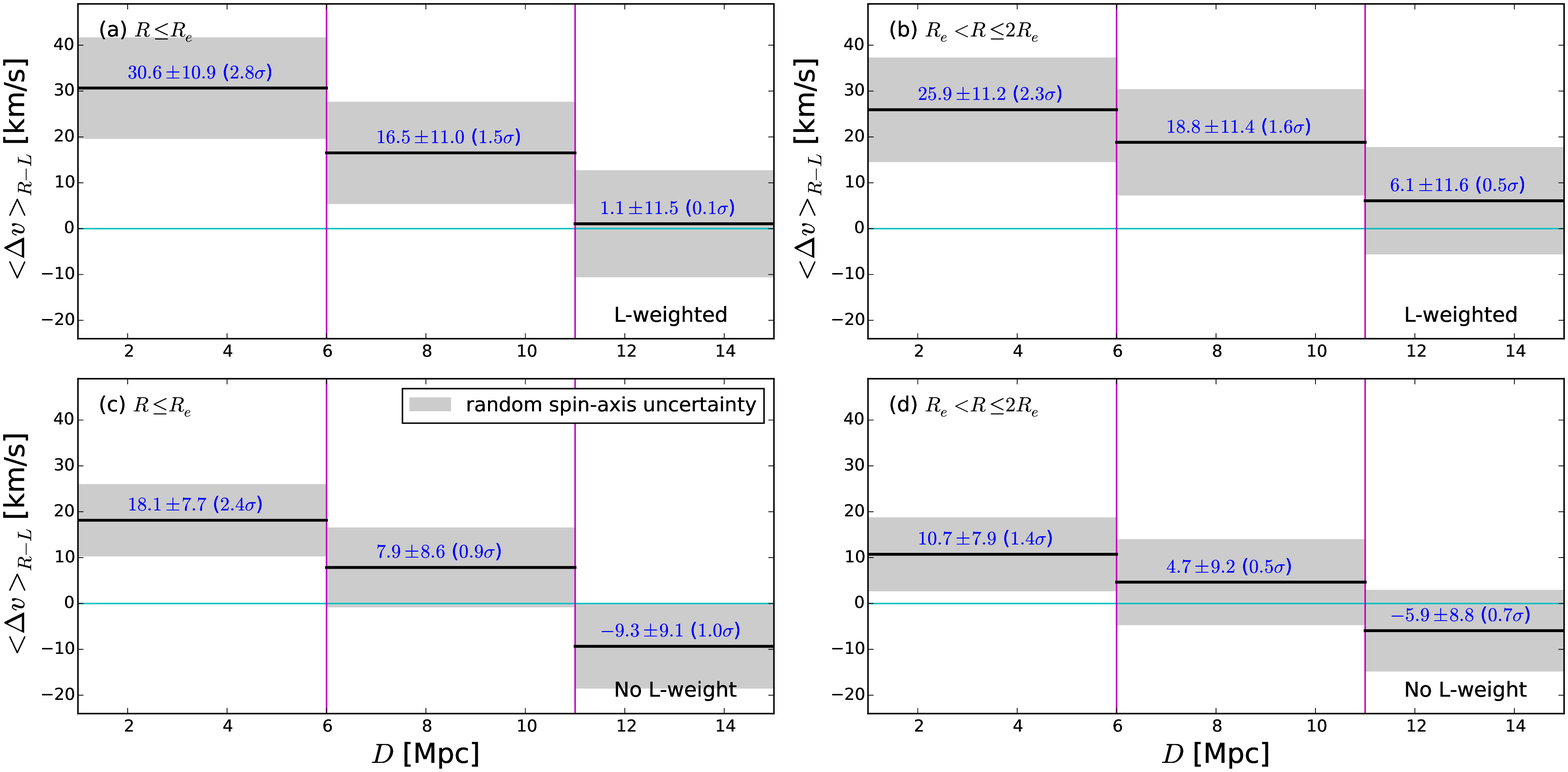}
\caption{Right-left-merged mean velocities at selected distance ranges for the whole CALIFA galaxies, when only red neighbors are used: (a) central rotation with luminosity weight, (b) outskirt rotation with luminosity weight, (c) central rotation without luminosity weight, and (d) outskirt rotation without luminosity weight. The random spin-axis (RAX) uncertainties are denoted (shades). \label{pltall}}
\end{figure*}

In \emph{L19}, two options for the luminosity weight ($\mathcal{L}$) were applied: absolute-luminosity weight and relative-luminosity weight. For the relative-luminosity weight, $\mathcal{L}$ must be the luminosity ratio between a neighbor and a given CALIFA galaxy, instead of the simple luminosity of a neighbor galaxy. The relative-luminosity-weighted mean velocity better reflects the direct interactions between a CALIFA galaxy and its neighbors. However, because such direct interactions are hardly expected in large (several Mpc) scales, we use only the absolute-luminosity-weighted mean velocities in this paper.

Note that the cumulative profiles (Figure~\ref{vprof}(b)) do not converge to zero velocity as $|D|$ increases but have small margins ($\sim25-45$ {\kms}) to positive direction (upward). This is probably due to the CALIFA target selection bias: the CALIFA targets are apparently much brighter than average NSA galaxies, which results in the tendency that the CALIFA targets are biased to lower redshifts compared to NSA galaxies (see Section~4.1 of L19 for more detailed discussion). We confirmed that the margin tends to be mitigated more when a tighter cut of apparent magnitude is applied to the neighbor galaxies, which strongly supports our interpretation.
Such margins are canceled out in the right-left merged profiles.
Figure~\ref{velmap}(b) shows the velocity contour map with the color bar shifted by 35 {\kms}, which gives us a clearer view for visually checking the existence of the dynamical coherence, by removing the bias-induced velocity-margin effect. 

In the finally-derived right-left-merged mean velocity profiles, the statistical uncertainty is estimated using three different methods: bootstrap (BST) uncertainty, random spin-axis (RAX) uncertainty, and randomly flipped ($\pm90^{\circ}$) spin-axis (RFA) uncertainty.
\begin{enumerate}
 \item[(5)] To estimate the BST uncertainty, the neighbors are randomly resampled with replacement, and the standard deviation of the resulting mean velocity profiles from 1000-times resampling experiments is estimated.
 \item[(6)] The estimation of the RAX uncertainty is based on a null hypothesis, ``the spin axis of each CALIFA galaxy is randomly determined regardless of the motions of its neighbors''. To test it, after replacing the angular momentum vector of each CALIFA galaxy with a random vector, we build a new (random-vector-based) composite kinematics map and derive its corresponding mean velocity profiles. The standard deviation is estimated from 1000-times repetition of this process.
 \item[(7)] The process to estimate the RFA uncertainty is similar to that for the RAX uncertainty, but the angular momentum vector of each CALIFA galaxy is randomly flipped by $+90^{\circ}$ or $-90^{\circ}$, instead of being fully randomized.
\end{enumerate}

In Figures~\ref{vprofall} - \ref{vprofneicol}, all the three kinds of uncertainty are presented at the same time, but we will regard the RAX uncertainty as the standard uncertainty. This is because the null hypothesis for the RAX test exactly coincides with what we intend to examine in this paper.
The BST uncertainty is classic and widely used, but it may vary according to the size of a sample, which tends to result in too large uncertainty at small $D$ or too small uncertainty at large $D$.
Actually, when compared to the {\emph L19} results, the BST uncertainty is not well matched at $D=1$ Mpc, whereas the RAX and RFA uncertainties show very good agreements.
The RFA uncertainty is useful to estimate genuinely random axis uncertainty, when it is assumed that there is some coherence between the CALIFA galaxy rotation and the motions of its neighbors, because the neighbors in the X-cut regions after the random flipping by $\pm90^{\circ}$ must have genuinely random motions, not contaminated by coherent motions. However, in the results, the difference between the RAX and RFA uncertainties appears to be tiny.

\section{RESULTS}\label{result}

\begin{figure}[t]
\centering
\plotone{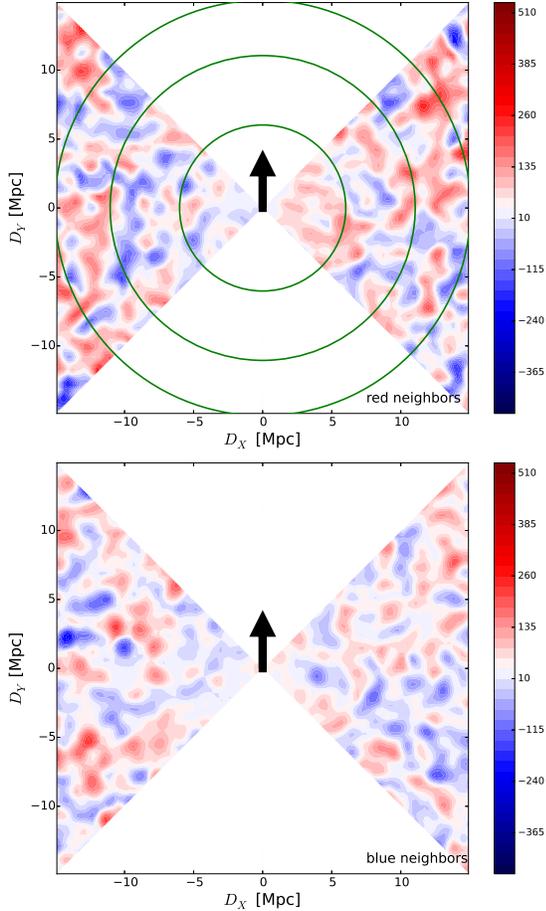}
\caption{\emph{Upper}: Contour map for the luminosity-weighted mean line-of-sight velocity of red ($g-r\ge0.7$) neighbors out to 15 Mpc (aligned for the central rotations of the CALIFA galaxies). The right-side bar shows the color code, in which the numbers indicate the line-of-sight velocities in unit of {\kms}. Note that the neutral point of the color bar indicates 35 {\kms}, not 0 {\kms} to compensate the redshift bias of the CALIFA galaxies. The three green circles show the distance ranges of 6, 11 and 15 Mpc, respectively. \emph{Lower}: Contour map for the luminosity-weighted mean line-of-sight velocity of blue ($g-r<0.7$) neighbors. \label{velmap2}}
\end{figure}

\begin{figure*}[t]
\centering
\includegraphics[width=0.95\textwidth]{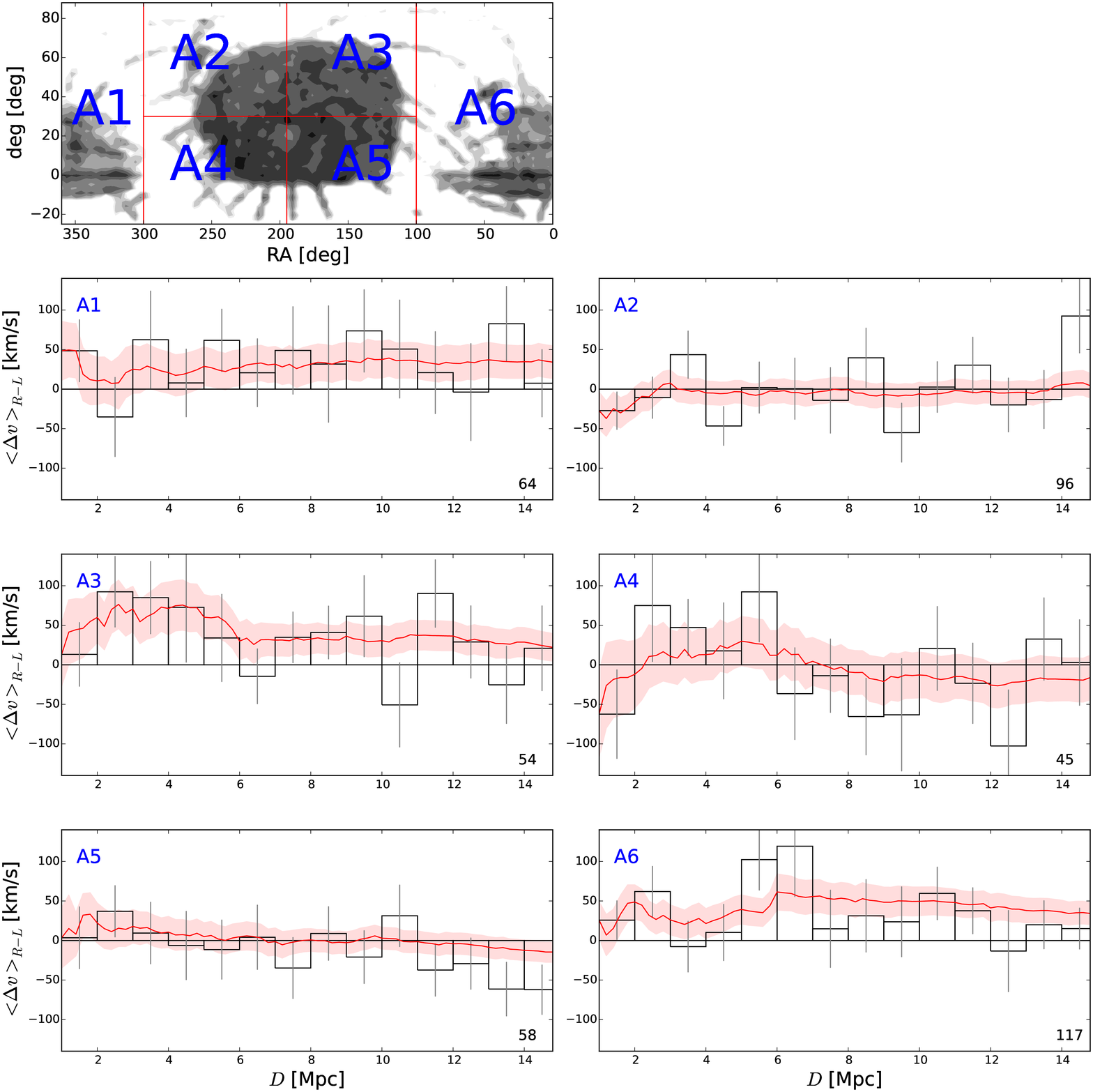}
\caption{The 1-Mpc-binned (histograms) and cumulative (red lines) luminosity-weighted mean velocity profiles for all CALIFA galaxies (central rotation) and red neighbors, when the sky is divided into six areas (A1 -- A6). The RAX uncertainties are overlaid (grey bars and red shades for the 1-Mpc-binned and cumulative velocities, respectively). The number of CALIFA galaxies in each area is denoted at the lower-right corner in each panel. \label{spfrag}}
\end{figure*}

In this section, the final products of the right-left-merged mean velocity profiles are inspected one by one. The results for the whole CALIFA sample and CALIFA subsamples divided by several quantities are presented in separate sub-sections.

\subsection{The Whole Sample}\label{result1}

Figure~\ref{vprofall} presents the 1-Mpc-binned mean velocity profiles and the cumulative mean velocity profiles for the whole sample of the CALIFA galaxies. 
In Figure~\ref{vprofall}(a) and (b), the binned mean velocities have positive values out to 8 Mpc, which is consistent with the coherent motion of neighbors aligned to the rotation of CALIFA galaxies. In the cumulative profiles, the coherence signal (i.e., the luminosity-weighted mean velocity of neighbors) is as large as $21.2\pm7.9$ {\kms} ($2.7\sigma$) at $D\le6.20$ Mpc for central rotation, while it is $22.1\pm8.4$ {\kms} ($2.6\sigma$) at $D\le2.95$ Mpc for outskirt rotation. The shapes of the cumulative profiles are possibly different between central and outskirt rotations (the cumulative mean velocity almost steadily increases out to 6 Mpc for central rotation, whereas the steady increase is only out to 3 Mpc for outskirt rotation), but the difference is statistically insignificant.
The significance to the BST uncertainty reaches to $2.9\sigma$ even at $D>10$ Mpc, but we will not overvalue it, because of the weakness of the BST uncertainties mentioned in Section~\ref{anal}. The RFA uncertainty tends to well follow the trends of the RAX uncertainty.

The properties of neighbors that have stronger coherent motions are important clues to infer the origin of this mysterious dynamical coherence in large scales. Thus, we estimated the mean velocity profiles for the whole CALIFA galaxies when their neighbors are controlled.
Figure~\ref{vprofneimag} compare the results when the neighbors are divided by luminosity ($M_r=-20$). In this comparison, the difference between bright and faint neighbors do not seem to be large, overall.

On the other hand, in Figure~\ref{vprofneicol}, the red and blue neighbors show striking differences. When the neighbors are limited to red ($g-r\geq 0.7$) galaxies, the coherence signals are as large as $29.1\pm9.7$ {\kms} ($3.0\sigma$) at $D\le6.20$ Mpc and $22.7\pm7.9$ {\kms} ($2.9\sigma$) at $D\le10.30$ Mpc (for central rotation; Figure~\ref{vprofneicol}(c)). 
The coherence signals for outskirt rotation are slightly smaller, but still considerable ($2.5-2.8\sigma$ significance; Figure~\ref{vprofneicol}(d)). These coherence signals are even more significant than those when the whole neighbors are used, despite the smaller neighbor sample size.
The binned mean velocities mostly have positive values (except the 8 - 9 Mpc bin for central rotation) out to 11 Mpc (Figure~\ref{vprofneicol}(a) and (b)), which supports the existence of dynamical coherence in such large scales, too.
On the other hand, when the neighbors are limited to blue ($g-r<0.7$) galaxies, coherence signals appear to be obviously insignificant. These results indicate that the color of the neighbors is a critical factor for the large-scale coherence.

\begin{deluxetable}{cr @{$\pm$} lc}
\tablenum{1} \tablecolumns{4} \tablecaption{Coherence Signal at $1 - 6$ Mpc in Each Sky Area} \tablewidth{0pt}
\tablehead{  Sky Area & \multicolumn{2}{c}{$\langle\Delta v\rangle_{(R-L)}\pm$ e$_{\textrm{\tiny RAX}}$} & Significance \\
& \multicolumn{2}{c}{[{\kms}]} }
\startdata
A1 & 22.1 & 27.2 & $0.8\sigma$ \\
A2 & 0.2 & 17.5 & $0.0\sigma$ \\
A3 & 31.2 & 26.7 & $1.2\sigma$ \\
A4 & 25.8 & 34.1 & $0.8\sigma$ \\
A5 & 5.0 & 19.0 & $0.3\sigma$ \\
A6 & 64.9 & 26.0 & $2.5\sigma$ \\
\hline\hline
RMS$^{\dagger}$ & \multicolumn{2}{l}{21.0}  \\
Error$^{\ddagger}$ & \multicolumn{2}{l}{\phantom{0}9.4}  \\
\enddata
\tablecomments{ $^\dagger$ The root-mean-square dispersion of $\langle\Delta v\rangle_{(R-L)}$ among the six sky areas. \\
$^\ddagger$ Error on the mean = RMS $/\sqrt{N-1}$.}
\label{skytab}
\end{deluxetable}

\begin{figure*}[t]
\centering
\includegraphics[width=0.95\textwidth]{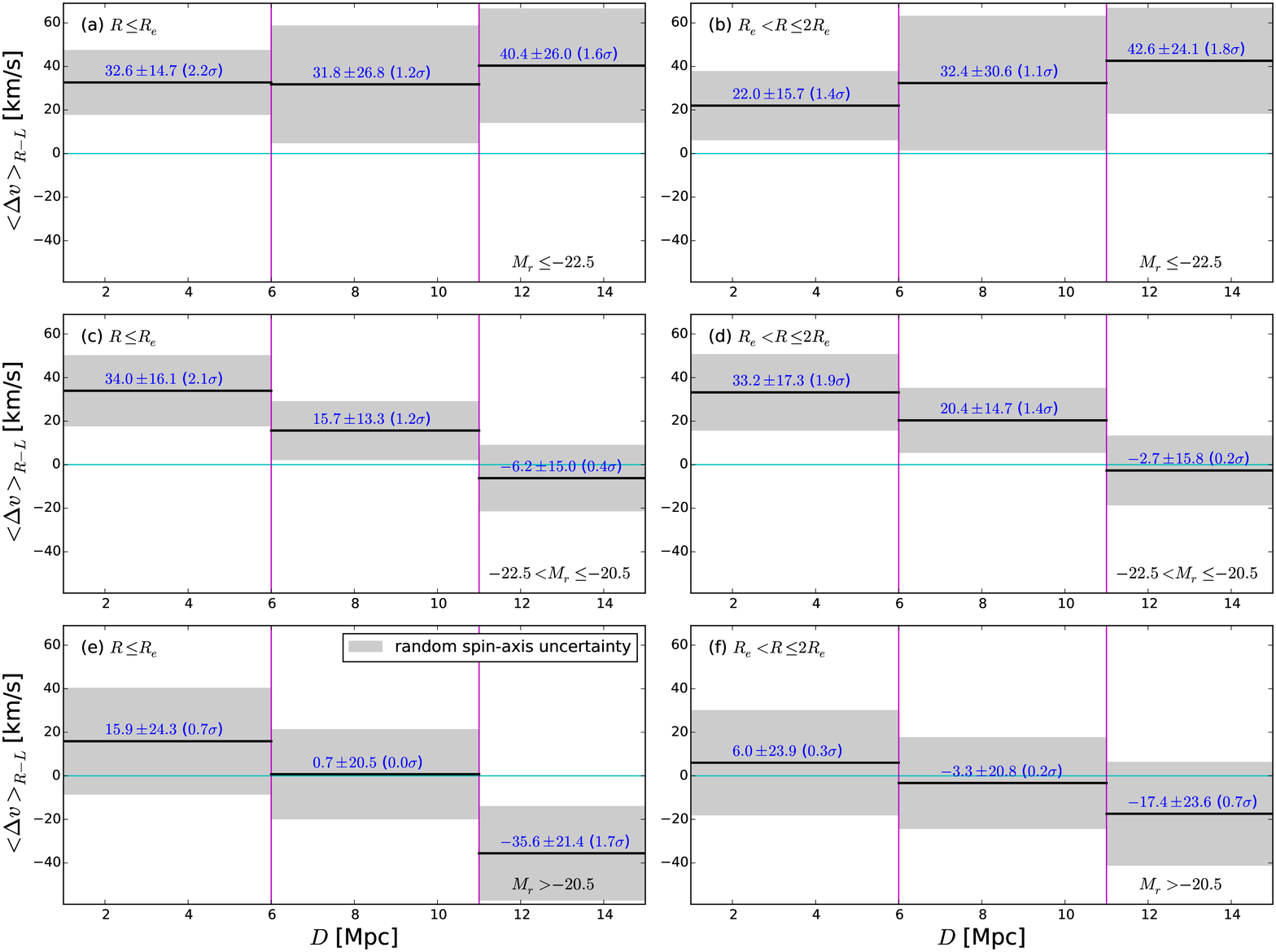}
\caption{Right-left-merged luminosity-weighted mean velocities at selected distance ranges for  the CALIFA subsamples divided by luminosity: (a) central rotation for bright ($M_r\leq -22.5$) galaxies, (b) outskirt rotation for bright galaxies, (c) central rotation for intermediate ($-22.5<M_r\leq -20.5$) galaxies, (d) outskirt rotation for intermediate galaxies, (e)  central rotation for faint ($M_r>-20.5$) galaxies, and (f) outskirt rotation for faint galaxies. Only red ($g-r\ge0.7$) neighbors are considered.\label{vprofmag}}
\end{figure*}

One important issue is the exact distance scale, out to which the dynamical coherence exists. When we focus on the central rotations and the red neighbors, the coherence out to 6 Mpc seems to be quite clear, because (1) all binned mean velocities have positive values, (2) the 5 - 6 Mpc bin shows the highest mean velocity (= the strongest coherence signal), and (3) the cumulative profile almost steadily increases. However, for the signals at 6 - 11 Mpc range, it is not easy to assert if the signals are sufficiently significant. The weakness of the cumulative profile is that once a strong signal appears (e.g., the high $\langle\Delta v\rangle_{R-L}$ at 5 - 6 Mpc), it may strongly influence the cumulative mean velocities even out of that point. In other words, the high coherence signals out to 11 Mpc in the cumulative profile may be simply the remnant effect of the strong coherence signals at $D\le6$ Mpc.

To address this issue, we plot Figure~\ref{pltall}, which shows the mean velocities at three selected distance ranges: $1 - 6$ Mpc, $6 - 11$ Mpc, and $11 - 15$ Mpc. In addition, we also compare the mean velocities with and without luminosity weight, to see how significantly the luminosity weight influence the results. As a result, we confirm that the coherence signal is still strong ($30.6\pm10.9$ {\kms}; $2.8\sigma$; for central rotation and with luminosity weight) at $D\le6$ Mpc, even after the influence of small-scale coherence ($<1$ Mpc) is removed. The significance becomes weaker when luminosity weight is not applied, but still meaningful ($18.1\pm7.7$ {\kms}; $2.4\sigma$). However, at the $6 - 11$ Mpc range, the statistical significance of dynamical coherence appears to be very marginal ($16.5\pm11.0$ {\kms}; $1.5\sigma$). That is, even though we suspect the existence of dynamical coherence out to 11 Mpc from Figure~\ref{vprofneicol} (consistently positive $\langle\Delta v\rangle_{R-L}$ out to 11 Mpc for red neighbors), the statistical evidence for it is not decisive. Thus, hereafter we will focus on the distance range of $D\le6$ Mpc, at which the obvious coherence signals are detected.

Figure~\ref{velmap2} presents the contour maps for the luminosity-weighted mean line-of-sight velocity of red and blue neighbors, respectively. Compared to Figure~\ref{velmap} (lower panel), the trends of `redshift at the right side' and `blueshift at the left side' appear more obviously when the neighbors are limited to red ones, particularly at $D\le6$ Mpc. The trends at $6 - 11$ Mpc range are somewhat ambiguous, and the $D>11$ Mpc range shows clearly no coherence signal.

Finally, we test if the large-scale coherence is a universal feature or there are some variations across the sky. Figure~\ref{spfrag} shows what the mean velocity profiles look like when the sky is divided into six areas. Although the coherence in each sky area is mostly insignificant because of the small sample size, the divided areas seem to present some possible differences: relatively strong coherence (A3 and A6), ambiguous coherence (A1 and A4), and almost no coherence (A2 and A5) out to 6 Mpc.
This may imply that the large-scale coherence is attributed to specific large-scale structures, rather than to a universal property in the Universe. We also estimated the luminosity-weighted mean velocity and its RAX uncertainty at the $1 - 6$ Mpc distance range in each sky area, the results of which are summarized in Table~\ref{skytab}. The root-mean-square (RMS) dispersion of the mean velocities among the six sky areas and the error on the mean (= RMS $/\sqrt{N-1}$) are also given. The RMS is comparable with the RAX uncertainty in each sky area, and the error of the mean estimated using the six sky subsamples gives a $3.3\sigma$ significance to the mean velocity of the whole sample ($30.6\pm9.4$ {\kms}).

In summary, the dynamical coherence is obviously detected out to 6 Mpc, with confidence levels up to $2.8\sigma$ significance. This is the first discovery of the dynamical coherence in such a large scale. We suspect the possible existence of dynamical coherence even out to 11 Mpc, but the statistical evidence is insufficient at least in this study.

\subsection{Subsamples}\label{result2}

\begin{figure*}[p]
\centering
\includegraphics[width=0.95\textwidth]{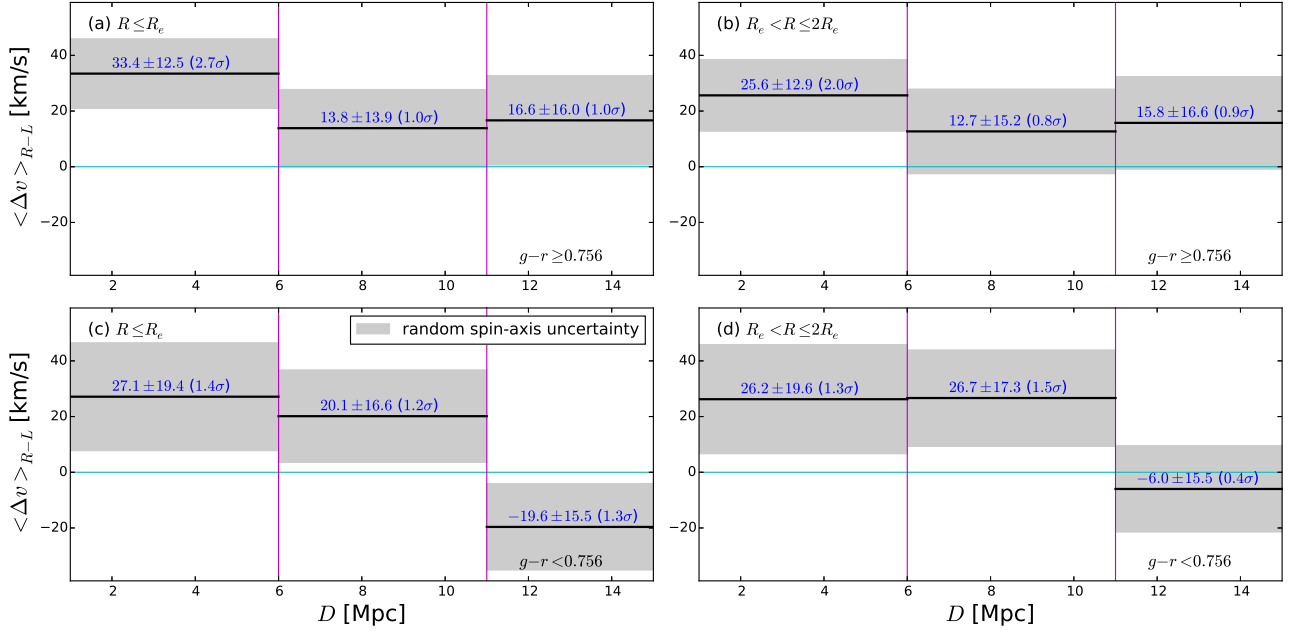}
\caption{Right-left-merged luminosity-weighted mean velocities at selected distance ranges for the CALIFA subsamples divided by color: (a) central rotation for red ($g-r\leq 0.756$) galaxies, (b) outskirt rotation for red galaxies, (c) central rotation for blue ($g-r>0.756$) galaxies, and (d) outskirt rotation for blue galaxies. Only red ($g-r\ge0.7$) neighbors are considered.\label{vprofcol}}
\end{figure*}

\begin{figure*}[p]
\centering
\includegraphics[width=0.95\textwidth]{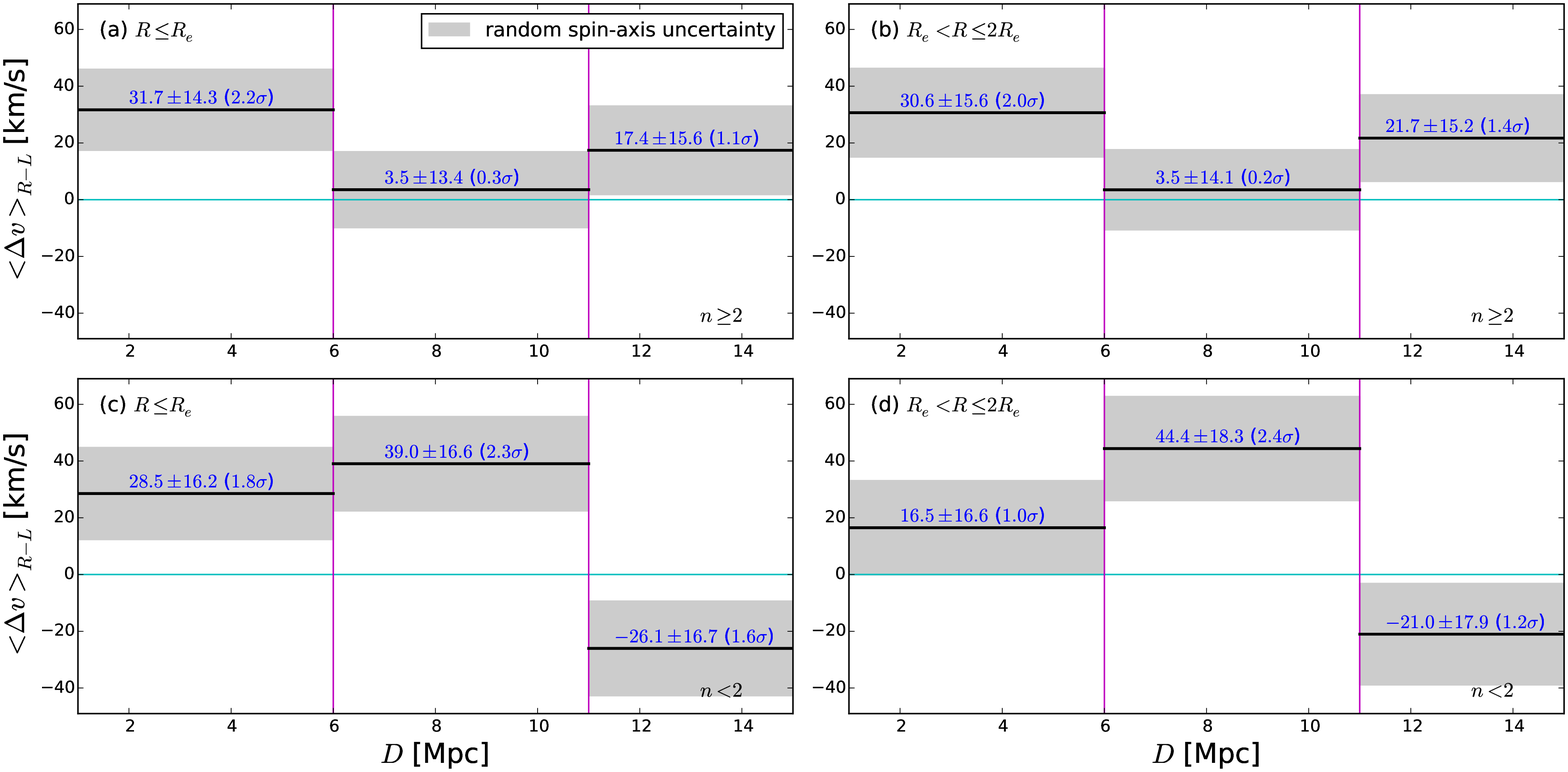}
\caption{Right-left-merged luminosity-weighted mean velocities at selected distance ranges for  the CALIFA subsamples divided by S{\'e}rsic index: (a) central rotation for concentrated ($n\leq 2$) galaxies, (b) outskirt rotation for concentrated galaxies, (c) central rotation for diffuse ($n>2$) galaxies, and (d) outskirt rotation for diffuse galaxies. Only red ($g-r\ge0.7$) neighbors are considered.\label{vprofser}}
\end{figure*}

\begin{figure*}[p]
\centering
\includegraphics[width=0.95\textwidth]{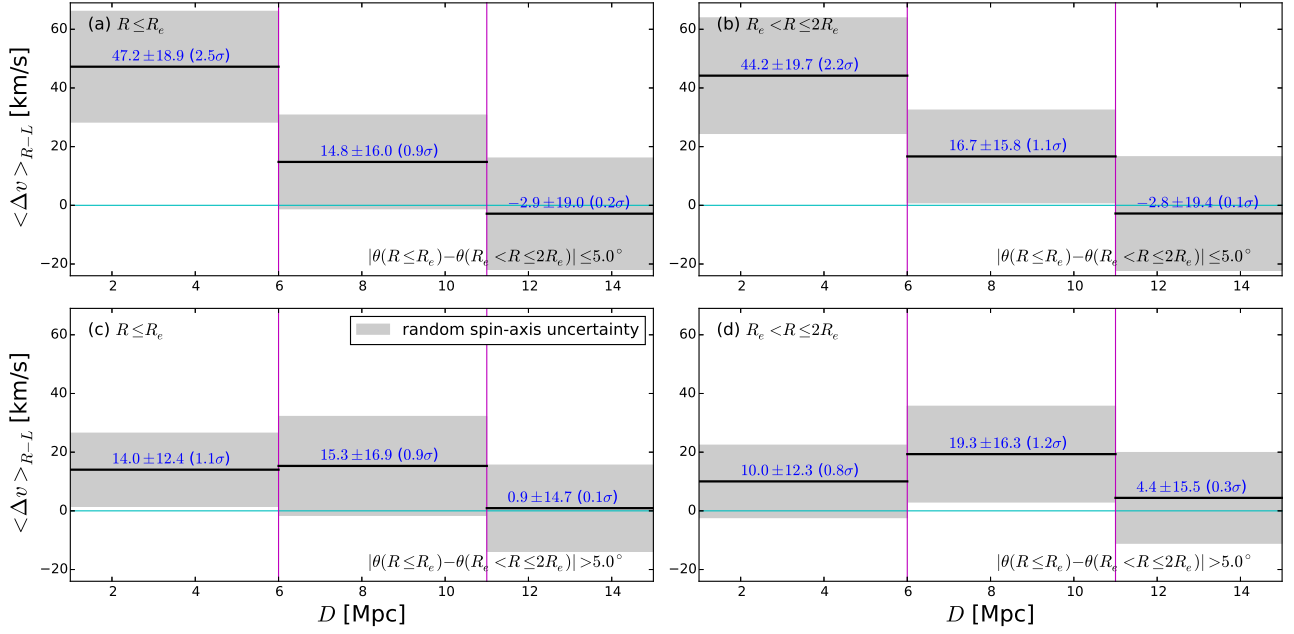}
\caption{Right-left-merged luminosity-weighted mean velocities at selected distance ranges for  the CALIFA subsamples divided by internal misalignment: (a) central rotation for well-aligned ($|\theta(R\le R_e)-\theta(R_e<R\le 2R_e)|\leq 5.0^{\circ}$) galaxies, (b) outskirt rotation for well-aligned galaxies, (c) central rotation for misaligned ($|\theta(R\le R_e)-\theta(R_e<R\le 2R_e)|>5.0^{\circ}$) galaxies, and (d) outskirt rotation for misaligned galaxies. Only red ($g-r\ge0.7$) neighbors are considered. \label{vprofmis}}
\end{figure*}

\begin{figure*}[p]
\centering
\includegraphics[width=0.95\textwidth]{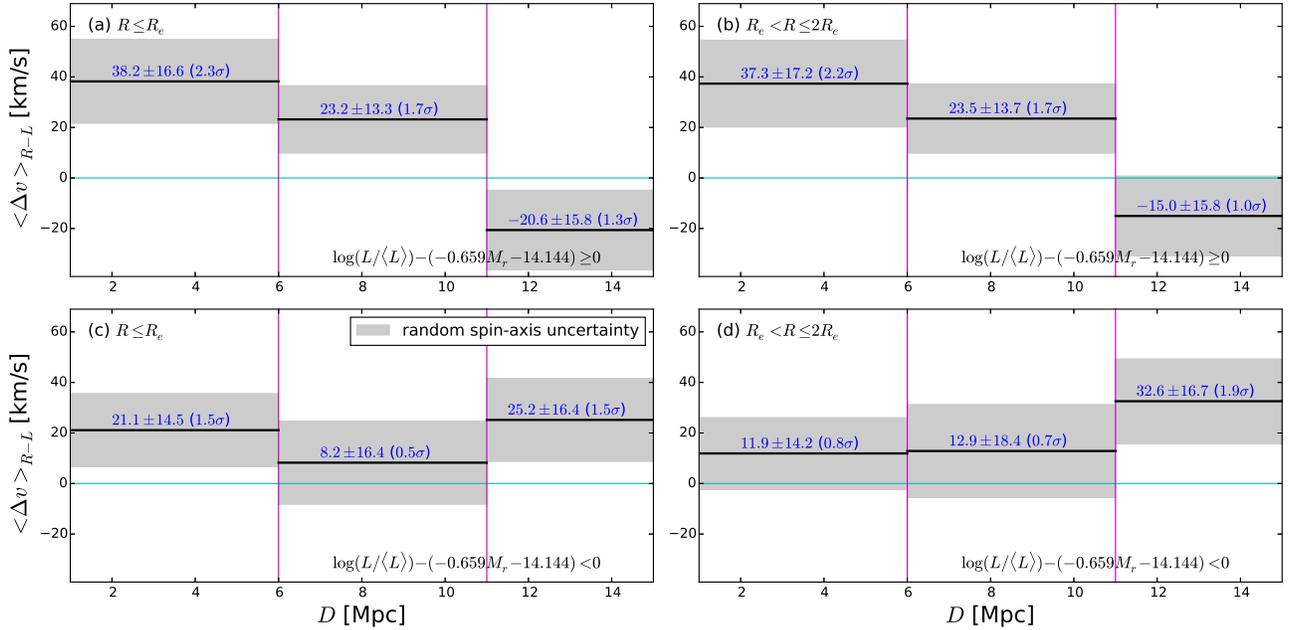}
\caption{Right-left-merged luminosity-weighted mean velocities at selected distance ranges for the CALIFA subsamples divided by luminosity-corrected central angular momentum: (a) central rotation for fast-rotating ($\log(L/\langle L\rangle)-(-0.659M_r - 14.144) \geq 0$) galaxies, (b) outskirt rotation for fast-rotating galaxies, (c) central rotation for slowly-rotating ($\log(L/\langle L\rangle)-(-0.659M_r - 14.144) < 0$) galaxies, and (d) outskirt rotation for slowly-rotating galaxies. Only red ($g-r\ge0.7$) neighbors are considered. \label{vprofmom}}
\end{figure*}

\begin{figure*}[t]
\centering
\includegraphics[width=0.95\textwidth]{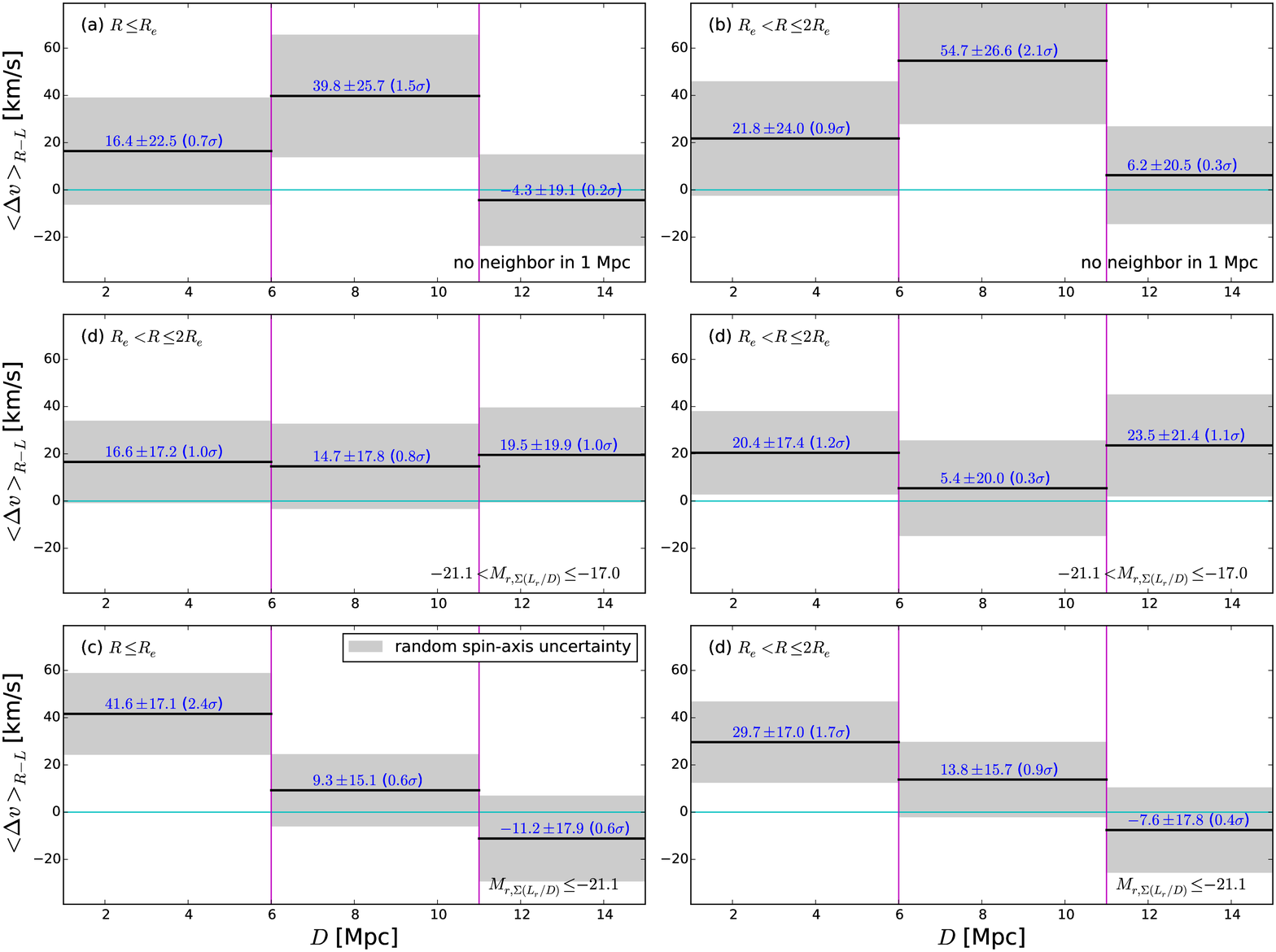}
\caption{Right-left-merged luminosity-weighted mean velocities at selected distance ranges for the CALIFA subsamples divided by local luminosity density: (a) central rotation for galaxies in loose environment (no neighbor with $M_r\leq -21$ in 1 Mpc), (b) outskirt rotation for galaxies in loose environment, (c) central rotation for galaxies in intermediate environment ($-21.1<M_{r,(\sum(L_r/D)}\leq -17.0$), (d) outskirt rotation for galaxies in intermediate environment, (e)  central rotation for galaxies in dense environment ($M_{r,(\sum(L_r/D)}\leq -21.1$), and (f) outskirt rotation for galaxies in dense environment. Only red ($g-r\ge0.7$) neighbors are considered. \label{vprofenv}}
\end{figure*}

We examine various subsamples of the CALIFA galaxies, divided by luminosity, color, S{\'e}rsic index, internal misalignment, luminosity-corrected angular momentum, and local luminosity density. In this subsample analysis, we limited the neighbors only to red ones, because the results in Section~\ref{result1} clearly show that the large-scale coherence is related to red neighbors, not blue ones. We tried these comparisons to find out any clues for the origin of the large-scale dynamical coherence. As a result, some cases show \emph{possible} differences between the subsamples, but unfortunately they are mostly statistically insignificant. 

Here we summarize the results of subsample comparisons. The statistical significance of the difference between the divided subsamples is given for each comparison, which is for central rotation and at $1-6$ Mpc if there is no additional note.
\begin{enumerate}
 \item[(1)] In Figure~\ref{vprofmag}, the bright CALIFA galaxies ($M_r\leq -20.5$) show relatively strong signals ($2.2 - 2.3\sigma$; at $1 - 6$ Mpc and for central rotation), whereas the faint CALIFA galaxies ($M_r> -20.5$) hardly show meaningful signals. [Significance of the difference: $0.6\sigma$]
 \item[(2)] In Figure~\ref{vprofcol}, the red ($g-r\ge0.756$) CALIFA galaxies show very slightly stronger coherence signals than blue ($g-r<0.756$) CALIFA galaxies ($2.7\sigma$ versus $2.2\sigma$). [Significance of the difference: $0.1\sigma$]
 \item[(3)] In Figure~\ref{vprofser}, both of the concentrated ($n\ge2$) and diffuse ($n<2$) galaxies show marginal coherence signals ($2.0 - 2.2\sigma$) at $1 - 6$ Mpc. However, at $6 - 11$ Mpc, the diffuse galaxies show marginal coherence signals ($2.3 - 2.4\sigma$) whereas the concentrated galaxies show no signal. This is the only case that the statistical significance of coherence signal is larger than $2\sigma$ at $6 - 11$ Mpc. [Significance of the difference: $1.7\sigma$ (at $6 - 11$ Mpc)]
 \item[(4)] In Figure~\ref{vprofmis}, the well-aligned galaxies are found to have stronger coherence signals ($2.5\sigma$) at $1 - 6$ Mpc, while the misaligned galaxies mostly show insignificant signals. [Significance of the difference: $1.5\sigma$]
 \item[(5)] In Figure~\ref{vprofmom}, the CALIFA galaxies with high angular momenta ($\log(L/\langle L\rangle)-(-0.659M_r - 14.144) \geq 0$) appear to be more strongly coherent with neighbors ($2.3\sigma$) than the ones with low angular momenta  at $1 - 6$ Mpc. [Significance of the difference: $0.8\sigma$]
 \item[(6)] In Figure~\ref{vprofenv}, the CALIFA galaxies in dense ($M_{r,(\sum(L_r/D)}\leq -21.1$) environment show stronger signals ($2.4\sigma$ at $1 - 6$ Mpc) than those in loose environment. [Significance of the difference: $1.0\sigma$]
\end{enumerate}

Again, we emphasize that these differences are \emph{statistically insignificant}. Only two cases show very marginal differences ($\gtrsim 1.5\sigma$): the diffuse or well-aligned CALIFA galaxies tend to show stronger signals of large-scale coherence. The other cases are too uncertain to be seriously discussed in this work, and we need to be sufficiently cautious even for the two very marginal cases.

\section{DISCUSSION}\label{discuss}

The key result in Section~\ref{result} is that galaxy rotation appears to be considerably coherent with the average line-of-sight motion of neighbors at far distances. When the neighbors are limited to red ones, the signal for the whole CALIFA sample is as significant as $2.8\sigma$ at $1<D\le$ 6 Mpc.
From this result, a simple but hard question is propounded.
\emph{How can the dynamical coherence be established over such large scales?}
Undoubtedly, direct interactions are impossible between galaxies separated by several Mpc. Then what caused this mysterious coherence in large scales?

The first clue is the property of the coherently-moving neighbors. In our results, only red neighbors show strong signals of dynamical coherence, while blue neighbors hardly show such signals. Red galaxies are widely used as a tracer of large-scale structures \citep[e.g.,][]{san09,kaz10,mon12,bau18}. In other words, the average motions of red neighbors may be equivalent with the motion of large-scale structures. If we adopt this interpretation, our results may indicate that the rotation of a galaxy is related to the motion of large-scale structures around it.

The second clue is the properties of the CALIFA galaxies with strong signals of large-scale coherence.
In Section~\ref{result2}, the diffuse or internally-well-aligned CALIFA galaxies tend to show stronger coherence signals, although the difference is very marginal. If we cautiously suppose that they are real features, such differences may be interpreted that late-type galaxies with less dynamical perturbation tend to have stronger large-scale coherence.
Hence, the two clues are combined into a single sentence: ``the rotational directions of late-type galaxies experiencing less dynamical perturbation are considerably related to the motions of large-scale structures around them''.

Before suggesting a scenario that explains this phenomenon, it will be worth comparing the results in this paper with those of \emph{L19}: the difference between the large-scale coherence and the small-scale coherence. In the small scale ($<$ 1 Mpc) of \emph{L19}, the rotations of \emph{faint} CALIFA galaxies are more strongly coherent with the average motion of \emph{bright} neighbors. On the other hand, in the large scale of this paper, the rotations of \emph{late-type} CALIFA galaxies show stronger coherence with the average motion of \emph{red} neighbors. About internal alignment of CALIFA galaxies, the small-scale coherence is stronger for \emph{misaligned} galaxies, whereas the large-scale coherence is stronger for \emph{well-aligned} galaxies. While all the features of small-scale coherence appear to be consistent with the interaction origin (\emph{L19}), the features of large-scale coherence found in this paper seem to be far from it. In other words, the two kinds of dynamical coherence probably have different origins.

One possible scenario for the large-scale dynamical coherence is as follows: A large-scale structure may have its own motion. The motion is different from the streaming motions of galaxies within the structure, but it indicates an extremely slow displacement of the structure itself. For example, imagine a large-scale filament or sheet with non-translational motion (different parts of the structure move at different speeds; differential motion).
If such a motion influences the individual angular momenta of the galaxies in the structure, then the large-scale dynamical coherence signals can manifest as discovered in this paper.

Unfortunately we do not have sufficient evidence supporting this scenario now, but we continue our speculation based on it. In our results, the luminosity-weighted mean velocity at $1<D\le6$ Mpc is 30.6 {\kms} (for central rotation of the CALIFA galaxies and for red neighbors; Figure~\ref{vprofneicol}). Supposing that this speed represents the long-term motion of large-scale structures (for example, the filament or sheet we assumed in the previous paragraph), we can roughly estimate the speed of position angle variation of the large-scale structure as follows:
30.6~{\kms} $\div$ 6 Mpc $\approx$ 2.9$^{\circ}$ per 10 Gyr.
Even if we adopt the luminosity-weighted mean velocity of 64.6 {\kms} for the A6 area (Table~\ref{skytab}), the speed of position angle variation is only 6.2$^{\circ}$ per 10 Gyr.
That is, in this speed, the change of the large-scale structure will be tiny even over the Hubble time.

If such a slow motion of a large-scale structure causes coherent angular momenta of galaxy-forming proto-clouds in it, the angular momenta will be conserved even after the proto-clouds form galaxies, until they suffer some disturbances from outside, such as galaxy interactions or merging events.
This scenario explains why unperturbed late-type galaxies show stronger coherence signals: late-type galaxies may conserve their initial angular momenta, whereas early-type galaxies grown through various merging events may have lost them.
The sky variation of the large-scale coherence found in Figure~\ref{spfrag} and Table~\ref{skytab} may be also explained by this scenario, because large-scale structures need to be well aligned perpendicularly to our line-of-sight, to be detected in our analysis.
However, we emphasize again that the differences between the subsamples are very marginal, and thus they need to be confirmed using a sufficiently large IFS sample, which will be crucial to support our suggested scenario.

How can we verify this scenario in another observational approach?
To do that, first it is necessary to (1) identify large-scale structures (such as filaments or sheets) that have a long-term motion as described above. After that, we need to (2) collect IFS data for a number of galaxies in the structures and (3) estimate the angular momentum vectors of those galaxies and compare their directions with the long-term motions of the structures. 
Since today various IFS surveys are producing data cubes for more and more galaxies, Steps (2) and (3) may not be too hard only if Step (1) is accomplished.

However, the real problem is Step (1): currently we cannot suggest any promising methodology to observationally confirm the long-term motion of a given large-scale structure. It is because the line-of-sight velocity of a large-scale structure (or the galaxies in it) is the combination of the Hubble expansion and the peculiar motion, which cannot be observationally distinguished. Thus, although a statistical study for a bundle of large-scale structures will be possible (just like this work), an intensive investigation for a given specific structure seems to be hardly achievable.
In that sense, numerical simulations would be a better approach practically, if it is possible that they are done for sufficiently large scales (to cover large-scale structures) and in high resolution (to resolve galaxy rotations) at the same time. %Otherwise, our results may be an important constraint on numerical simulations to improve their reality.

Finally, we try to reconcile this scenario with the previous findings that the spin axes of galaxies are aligned with large-scale filaments. According to recent studies in simulations \citep{nav04,ara07,bru07,cen14,dub14,liu17,lee18} and in observations \citep{tem13,zha13,zha15,hir17,kim18,jeo19},
late-type galaxies in a filament tend to have spin axes parallel with the filament direction, while spin axes of early-type galaxies tend to be perpendicular to it.
Since the galaxies with strong coherence signals in our results may be mainly late-type galaxies, if they are located in filaments, their spin axes may be aligned to be parallel with filaments according to those studies.
In this case, it is not strongly expected that a late-type galaxy in a filament has large-scale dynamical coherence with galaxies in the same filaments, even if the filament has its own long-term differential motion. However, suppose that the filament is embedded in a sheet-like structure with its own long-term differential motion, and this motion had induced the spin of the late-type galaxy. Then, the late-type galaxy in the filament will have the large-scale dynamical coherence with sheet galaxies, rather than with other filament galaxies. In this way, our scenario and the previous studies can be reconciled. As mentioned earlier, such a configuration of large-scale structures can not be easily identified in observations. However, with the help of simulations, it could be explored more along this direction.

\section{CONCLUSION}\label{conclude}

We examined whether there is any coherence between the rotational direction of galaxies and the average motions of their neighbor galaxies in large scales out to 15 Mpc, using the CALIFA survey data and the NSA catalog.
From our statistical analysis, we discovered that the coherence is established even in several-Mpc scales.
Our main conclusions are summarized as follows:
\begin{enumerate}
 \item[1.] The rotation of a galaxy appears to be related to the average motion of its neighbors out to several Mpc scales. The large-scale coherence is stronger when the neighbors are limited to red ones ($2.8\sigma$ significance at $1<D\le6$ Mpc for central rotation), whereas it is obviously insignificant for blue neighbors.
 \item[2.] The diffuse or internally-well-aligned CALIFA galaxies show stronger coherence signals than concentrated or internally-misaligned CALIFA galaxies. However, the differences are statistically very marginal and thus need to be checked using a much larger IFS sample.
 \item[3.] The detailed trends of the large-scale coherence are different from those of the small-scale coherence. The features of the large-scale coherence seem to be hardly caused by direct interactions between galaxies, which were suggested as the main origin of the small-scale coherence in \emph{L19}.
 \item[4.] For the large-scale coherence discovered in this paper, we cautiously suggest a scenario that the long-term motion of a large-scale structure may influence the rotations of galaxies in it. It will not be easy to verify this scenario in another observational approach, but numerical simulations would be helpful.
\end{enumerate}

\acknowledgments
This study uses data provided by the Calar Alto Legacy Integral Field Area (CALIFA) survey (http://califa.caha.es/), which is based on observations collected at the Centro Astron{\'o}mico Hispano Alem{\'a}n (CAHA) at Calar Alto, operated jointly by the Max-Planck-Institut f{\"u}r Astronomie and the Instituto de Astrof{\'i}sica de Andaluc{\'i}a (CSIC).
This study also uses the the NASA/IPAC Extragalactic Database (NED), which is operated by the Jet Propulsion Laboratory, California Institute of Technology, under contract with the National Aeronautics and Space Administration.

\end{document}